\documentclass[aps,pra,superscriptaddress]{revtex4-2}

\usepackage{graphicx}
\usepackage{amsmath,amssymb,amsfonts}
\usepackage{bm}
\usepackage{hyperref}
\usepackage{titlesec}

\begin{document}

\titleformat{\section}
{\normalfont\large\bfseries}
{\arabic{section}.}
{1em}
{}

\renewcommand{\thesubsection}{(\alph{subsection})}
\titleformat{\subsection}
{\normalfont\normalsize\bfseries}
{(\alph{subsection})}
{0.8em}
{}

\makeatletter
\renewcommand{\p@subsection}{}
\renewcommand{\p@subsubsection}{}
\makeatother

\title{Further fundamental bounds on the Hall effect\\ in three-dimensional metamaterials}

\author{Christian Kern}
\email{physics@chrkern.de}
\affiliation{Department of Civil and Mechanical Engineering, Technical University of Denmark, 2800 Kgs. Lyngby, Denmark}
\author{Graeme W. Milton}
\email{graeme.milton@utah.edu}
\affiliation{Department of Mathematics, University of Utah, Salt Lake City, UT 84112, USA}

\begin{abstract}
We consider the problem of bounding the effective nonreciprocal properties of metamaterials. Recently, significant progress was made by showing that this problem can be reduced to bounding an equivalent reciprocal one and applying a monotonicity argument. Here, we build upon this result and provide bounds incorporating additional information about the metamaterial. Specifically, in the two-phase case, we provide bounds for isotropic metamaterials. In the multiphase case, we provide bounds for uniaxial metamaterials that additionally incorporate the volume fractions. In both instances, the incorporated additional information significantly tightens the bounds. In the two-phase case, we evaluate the bounds by comparing them to the effective properties of hierarchical laminate microstructures. This comparison ultimately leads us to propose a set of conjectured bounds. While our discussion focuses on the Hall effect, our results are more broadly applicable to other nonreciprocal effects in so far as their mathematical description is equivalent. In particular, our bounds apply to the Faraday effect in the quasistatic regime and in the absence of losses and resonances.
\end{abstract}

\maketitle

\section{Introduction}

\noindent
A key problem in the theory of metamaterials is to characterize the range of the effective properties as the metamaterial's geometry is varied \cite{Milton:2002:TOC}. This characterization entails both deriving (outer) bounds on the effective properties and showing that these bounds are attained by microstructures. Historically, this problem has been much studied in the reciprocal case, in which the material tensor is symmetric, as implied by Onsager's principle \cite{Onsager:1931:RRI,Onsager:1931:RRII,Landau:1980:STP:120}. 
\smallskip

When an external bias breaks time-reversal symmetry, materials can exhibit nonreciprocal behavior, which is reflected in an antisymmetric contribution to the material tensor \cite{Landau:1980:ECM:22}. The canonical example is the application of an external magnetic field, which gives rise to, inter alia, the Hall effect \cite{Hall:1879:MEC,Landau:1980:ECM:22} and the Faraday effect \cite{Landau:1980:ECM:12}. In the past, substantial progress has been made on characterizing the range of effective Hall properties of two-dimensional metamaterials \cite{Dykhne:1970:APR,Shklovskii:1976:CBH,Stroud:1984:NER,Milton:1988:CHE,Briane:2008:HTD}. For three-dimensional nonreciprocal metamaterials, it has been shown that a particularly rich behavior can be obtained, including the parallel Hall effect \cite{Briane:2010:AEH,Kern:2015:PHE,Kern:2017:EPH}, sign-inversions of the effective Hall coefficient \cite{Briane:2009:HTD,Kadic:2015:HES,Kern:2016:EES,Kern:2019:HBD}, and strong anisotropies of the magnetoresistivity \cite{Bergman:1994:CSF,Tornow:1996:AMC,Bergman:1997:RAM}. While corresponding bounds have been derived and significantly advanced our understanding \cite{Briane:2009:GHE,Briane:2011:BSF}, tight bounds have remained elusive until recently.
\smallskip

In a recent paper \cite{Kern:2025:BHE}, we have shown that tight bounds on nonreciprocal effective properties of three-dimensional metamaterials can be derived by exploiting the monotonicity of the effective tensor. In essence, one can instead bound the effective properties of a fictitious reciprocal metamaterial and relate them to the effective properties of the nonreciprocal metamaterial through a monotonicity argument. This approach has yielded comprehensive bounds and key implications, including the impossibility of enhancing the effective Hall mobility beyond that of constituent materials. However, several important cases have remained unaddressed.
\smallskip

In this paper, we extend the existing approach by providing specialized bounds that incorporate additional microstructural information. In the two-phase case, we derive bounds for isotropic metamaterials, which have attracted significant interest \cite{Miller:2017:FHE,Notomi:2017:RHE}. Both volume-fraction dependent and independent bounds are given and conclusions for the range of the effective Hall coefficient are drawn. To assess the tightness of the bounds, we examine hierarchical laminates, which are known to exhibit optimal performance in many instances \cite{Milton:2002:TOC}, including uniaxial Hall metamaterials \cite{Kern:2025:BHE}. While one point on the bounds is attained by a hierarchical laminate, substantial gaps remain, which leads us to conjecture a set of improved bounds. 
\smallskip

For multi-phase uniaxial metamaterials, we develop bounds that incorporate the volume fractions of the phases. Depending on the chosen parameter values, these bounds are significantly tighter than their volume-fraction independent counterpart \cite{Kern:2025:BHE}. Thus, if the volume fractions of the phases are known, our new bounds give a much clearer picture of the effective Hall properties. Furthermore, we show that at least one point on these bounds is attained by a microstructure, specifically by a rank-$1$ laminate.
\smallskip

Throughout the paper, we assume that the magnetic field and, thus, the antisymmetric contribution to the material tensor is small. Beyond the Hall effect, our results apply to other nonreciprocal effects to the extent that the mathematical description is equivalent, see also Ref.\,\onlinecite{Kern:2025:BHE}. In particular, in the absence of losses and resonance, they apply to the Faraday effect in the quasistatic regime.
\smallskip

The rest of this paper is organized as follows: In Sec.\,2, we outline the mathematical framework and strategy underlying the derivation of the bounds. Sec.\,3 provides the two-phase bounds for isotropic metamaterials and discusses their attainability as well as a set of conjectured improved bounds. In Sec.\,4, we present and discuss the multi-phase bounds for uniaxial metamaterials, before we conclude and give a brief outlook in Sec.\,5.  

\section{Preliminaries}

\noindent
In this section, we will briefly introduce the mathematical description of the Hall effect in metamaterials. Further details can be found in Refs.\,\onlinecite{Kern:2018:THE,Kern:2023:SRE}, Chaps.\,2.2 and 16.3 of Ref.\,\onlinecite{Milton:2002:TOC} and, from a mathematically rigorous perspective using the notion of $H$-convergence \cite{Murat:1997:C}, in Ref.\,\onlinecite{Briane:2009:HTD}. Subsequently, we will summarize the strategy for deriving bounds on the effective Hall parameters put forward in Ref.\,\onlinecite{Kern:2025:BHE}, which forms the basis of our analysis here.

\subsection{Mathematical description}

\noindent
In the presence of a magnetic field, $\bm{B}$, electric conduction is governed by the equations
\begin{equation}
	\nabla\cdot\widetilde{\bm{j}}=0, ~\nabla\times\widetilde{\bm{e}}=0, ~\widetilde{\bm{j}}=\widetilde{\bm{\sigma}}(\bm{B})\widetilde{\bm{e}},
	\label{eq:conductivityprob}
\end{equation}
where $\widetilde{\bm{j}}$ and $\widetilde{\bm{e}}$ are the current density and electric field, respectively, and $\widetilde{\bm{\sigma}}(\bm{B})$ is the magnetic-field dependent conductivity tensor. The tilde denotes that these quantities describe conduction in the presence of a magnetic field -- for the corresponding zero-magnetic field (zmf) quantities, we omit the tilde. We assume that the magnetic field is weak and only consider terms up to the first order in $\bm{B}$. Under this assumption, the constitutive law reads
\begin{equation}
	\widetilde{\bm{j}}=\bm{\sigma}\widetilde{\bm{e}}+(\bm{S}\bm{B})\times\widetilde{\bm{e}},
	\label{eq:constrel}
\end{equation}
where $\bm{\sigma}$ is the zmf conductivity tensor and $\bm{S}$ is a rank-two tensor that characterizes the first-order contribution to the magnetic-field dependent conductivity. Note that the same equations describe the Faraday effect in magneto-active materials in the quasistatic regime with the electric displacement field, the zmf permittivity tensor, and the magnetogyration tensor playing the role of the electric current density, the zmf conductivity, and the magneto-gyration tensor, respectively.
\smallskip

In some instances, instead of considering the $S$-tensor, $\bm{S}$, one prefers to work with the Hall tensor, $\bm{A}$, which can be regarded as the corresponding inverted quantity, i.e., it plays the corresponding role in the inverted version of the constitutive law (again only considering terms up to the first order),
\begin{equation}
	\widetilde{\bm{e}}=\bm{\rho}\widetilde{\bm{j}}+(\bm{A}\bm{B})\times\widetilde{\bm{j}},
\end{equation}
where $\bm{\rho}=\bm{\sigma}^{-1}$ is the resistivity tensor. One obtains the relation
\begin{equation}
	\bm{S} = -\text{Cof}\left(\bm{\sigma}\right)\bm{A},
	\label{eq:cof}
\end{equation}
where $\text{Cof}(\cdot)$ is the cofactor matrix. Note that the Hall tensor, $\bm{A}$, can be seen as the generalization of the Hall coefficient, $A$, to anisotropic materials, i.e., for isotropic materials, we have $\bm{A}=A\bm{I}$.
\smallskip

In a metamaterial, we distinguish between the fields and material moduli on a microscopic and on a macroscopic length scale. On the microscopic length scale, the behavior is governed by Eqs. (\ref{eq:conductivityprob}) and (\ref{eq:constrel}). On the macroscopic length scale, the metamaterial acts like a homogeneous material with certain ``effective'' properties -- here, an effective zmf conductivity tensor, $\bm{\sigma}^*$, and an effective $S$-tensor, $\bm{S}^*$, which relate the averaged fields on the macroscopic scale,
\begin{equation}
	\langle\widetilde{\bm{j}}\rangle=\bm{\sigma}^*\langle\widetilde{\bm{e}}\rangle+(\bm{S}^*\bm{B})\times\langle\widetilde{\bm{e}}\rangle.
\end{equation}
Correspondingly, one can define an effective Hall tensor, $\bm{A}^*$, which is related to the effective $S$-tensor via the macroscopic version of Eq.\,(\ref{eq:cof}).
\smallskip

A key goal of our paper is to derive (outer) bounds on the effective properties, i.e., bounds constraining (and relating) the effective zmf conductivity tensor and the effective $S$-tensor, as well as derived quantities. We assume that the metamaterial is made from isotropic phases, whose properties we denote as $\sigma_i$, $S_i$  $(i=1,\dots,m)$. The magnetic-field dependent conductivity tensor of each of the phases thus takes the form 
\begin{equation}
	\widetilde{\bm{\sigma}}_i=\begin{pmatrix} \sigma_i && -\eta_i && 0 \\ \eta_i && \sigma_i && 0 \\ 0 && 0 && \sigma_i \end{pmatrix} \text{ with } \eta_i = S_iB,
\end{equation}
assuming that the magnetic field is pointing along the $x_3$-direction. In the first part of the paper, we present bounds for (two-phase, $m=2$) isotropic metamaterials, i.e., $\bm{\sigma}^*=\sigma^*I$ and $\bm{S}^*=S^*\bm{I}$. In the second part of the paper, we present bounds for (multi-phase, $m\geq 2$) uniaxial metamaterials, i.e., 
$\bm{\sigma}^*=(\sigma_{\perp}^*,\,\sigma_{\perp}^*,\sigma_{\parallel}^*)$ and $\bm{S}^*=(S_{\perp}^*,\,S_{\perp}^*,S_{\parallel}^*)$,
if we choose the coordinate system such that the $x_3$-axis aligns with the uniaxial axis of symmetry. The corresponding magnetic-field dependent conductivity tensor is given by
\begin{equation}
	\widetilde{\bm{\sigma}}^*=\begin{pmatrix} \sigma_{\perp}^* && -\eta^* && 0 \\ \eta^* && \sigma_{\perp}^* && 0 \\ 0 && 0 && \sigma_{\parallel}^* \end{pmatrix} \text{ with } \eta^* = S_{\parallel}^*B,
\end{equation}
Note that the uniaxial bounds are subsequently extended to fully anisotropic metamaterials.
\smallskip

Before introducing the strategy for deriving our bounds in the next section, we briefly summarize how the effective Hall properties of metamaterials, including those that we will discuss in this paper, can be determined. The effective zmf conductivity tensor is given by
\begin{equation}
	\bm{\sigma}^*=\langle\bm{\sigma}\bm{E}\rangle,
\end{equation}
where $\bm{E}$ is the ``matrix-valued electric field'' (also referred to as the ``corrector'' \cite{Murat:1997:C,Briane:2009:HTD}) and the average is taken over the unit cell of periodicity. The columns of $\bm{E}$ are three microscopic electric fields satisfying the zero magnetic-field conductivity problem and being normalized, $\langle\bm{E}\rangle=\bm{I}$. If the magnetic field is small, it suffices to know $\bm{E}$, i.e. the microscopic electric field in the \textit{absence} of a magnetic field, to determine $\bm{S}^*$. Specifically, building upon a previous result by Bergman \cite{Bergman:1983:SDL}, Briane and Milton \cite{Briane:2009:HTD} derived the perturbation formula 
\begin{equation}
	\bm{S}^{*}=\langle\text{Cof}(\bm{E})^{\intercal}\bm{S}\rangle.
	\label{eq:effective_S}
\end{equation}
For laminates, i.e., microstructures resulting from a (repeated) layering process, the microscopic electric field follows from elementary considerations, see, e.g., Chap.\,9 in Ref.\,\onlinecite{Milton:2002:TOC}.

\subsection{Strategy for deriving the bounds}\label{derstrategy}

\noindent
The key insight in Ref.\,\onlinecite{Kern:2025:BHE} is that bounds for nonreciprocal metamaterials are implied by bounds for corresponding fictitious reciprocal metamaterials through a monotonicity argument. First, note that we can assume that the $\widetilde{\bm{\sigma}}_i$ are Hermitian, as $\widetilde{\bm{\sigma}}^*$ is an analytic function and the magnetic field is small. This corresponds to taking the magnetic field to be purely imaginary. (A related strategy was used by Schulgasser and Hashin to obtain bounds on the complex permittivity in the small-loss limit \cite{Schulgasser:1976:BEP}.) Then, if the conductivity tensors of the phases are increased from $\widetilde{\bm{\sigma}}_i$ to $\bm{\sigma}_i^+= (\sigma_i+|\eta_i|)\bm{I}$, where the $\bm{\sigma}_i^+$ are the conductivity tensors of the phases of the fictitious metamaterial, the monotonicity of the effective tensor implies that \cite{Kern:2025:BHE}
\begin{equation}
	\sigma^{*+} \geq \sigma^*+|\eta^*| \text{ and } \sigma_{\perp}^{*+} \geq \sigma_{\perp}^*+|\eta^*|,
	\label{eq:estimatebound}
\end{equation}
for isotropic and uniaxial metamaterials, respectively. Thus, an upper bound on $\sigma^{*+}$ (or $\sigma_{\perp}^{*+}$) implies a bound on $|\eta^*|$. Preferentially, the upper bound for the fictitious reciprocal metamaterial should also take into account the known value $\sigma^{*+}(\eta_i=0)=\sigma^*$ (or $\sigma_{\perp}^{*+}(\eta_i=0)=\sigma_{\perp}^*$). Our two-phase analysis is based on such bounds obtained through the field-equation recursion method starting from previous bounds by Bergman \cite{Bergman:1976:VBS,Kern:2025:BHE}. In the multi-phase case, we will generalize an approach by Prager \cite{Prager:1969:IVB} to obtain suitable bounds incorporating a known value. Note that, as we assume that the $\eta$-coefficients are small, it suffices to consider bounds correlating the effective conductivity with its derivative. As an alternative to the path followed here, such bounds can be derived from bounds on the effective complex permittivity using the argument of Schulgasser and Hashin put forward in Ref.\,\onlinecite{Schulgasser:1976:BEP} in an inverse fashion. Indeed, following this recipe, the bound on the effective complex permittivity corresponding to microstructures having a single resonance/pole, Eq.\,(24) in Ref.\,\onlinecite{Milton:1981:BCP}, yields the upper bound of our two-phase bounds (see Sec.\,3(a)) in an alternative way. Finally, we remark that, in a more general context, monotonicity arguments are well established in the theory of metamaterials, see, e.g., Ref.\,\onlinecite{Tartar:1979:ECH} and Chap.\,13.2 in Ref.\,\onlinecite{Milton:2002:TOC}.
\smallskip

The bounds are further improved through translations. In the simplest case, a translation is the addition of a constant offset to the microscopic conductivity tensor if it causes the effective conductivity tensor to shift by the same offset. Specifically, we will use that if we shift the $\eta$-coefficients by a constant $\eta_i \rightarrow \eta_i -\lambda$, then the effective $\eta$-coefficient will shift by the same constant, $\eta^* \rightarrow \eta^* -\lambda$ \cite{Stroud:1984:NER}. Note that, to apply the bounds, the antisymmetric part of the conductivity tensor has to remain small under a translation. If the translation parameter, $\lambda$, is chosen carefully, one can often obtain substantially tighter bounds. For the two-phase case, as in our previous paper \cite{Kern:2025:BHE}, we set $S_2=0$ without loss of generality by appropriate choice of $\lambda$. For the multiphase case, the optimal value of the translation parameter $\lambda$ is determined numerically.

\section{Bounds for two-phase isotropic metamaterials}

\noindent
In this section, we provide bounds applying to isotropic metamaterials made from two isotropic phases. The phases have conductivities $\sigma_1$ and $\sigma_2$, which we label such that $\sigma_1 > \sigma_2$, and $S$-coefficients $S_1$ and $S_2$. To simplify the discussion, we set $S_2=0$. Note that this does not limit the generality of the discussion, as we can always obtain the corresponding result for $S_2\neq 0$ via a translation. Furthermore, we will be expressing the effective properties in terms of the so-called $y$-tensor, which in our case takes the form 
\begin{align}
	\widetilde{\bm{Y}} &=\begin{pmatrix} y_{\sigma} & -y_{S} & 0 \\ 
		y_{S} & y_{\sigma} & 0 \\
		0 & 0 & y_{\sigma}
	\end{pmatrix}
	\quad \text{with}\\
	\widetilde{y}=y_{\sigma}+iy_{S} &=-f_2(\sigma_{1}+iS_1B)-f_1\sigma_{2}+\frac{f_1f_2(\sigma_{1}+iS_1B-\sigma_{2})^2}{f_1(\sigma_{1}+iS_1B)+f_2\sigma_{2}-(\sigma^*+iS^*B)},
	\label{eq:ytransf}
\end{align} 
where $i$ is the imaginary unit \cite{Briane:2011:BSF}. Using the $y$-tensor has the advantage of simplifying the treatment substantially. In particular, when expressed in terms of the $y$-tensor, the volume-fraction dependence of the bounds and the effective properties of many of the extremal microstructures is hidden in the transformation (\ref{eq:ytransf}). Beyond this convenient property, the $y$-tensor has physical meaning: While the effective conductivity tensor relates the averages of the fields, the $y$-tensor relates the average fluctuations of the fields in either phase. Our derivation of the bounds below will make use of the field-equation recursion method, which is based on considering a sequence of effective tensors corresponding to sequentially higher-order fluctuations with the $y$-tensor being the second tensor in this sequence. For further details see Refs.\,\onlinecite{Milton:1987:MCEa,Milton:1987:MCEb,Milton:1991:FER,Clark:1997:CFR,Milton:1985:TCC} and Chap.\,29 in Ref.\,\onlinecite{Milton:2002:TOC}. 
\smallskip

The obtained volume-fraction dependent bounds improve on the uniaxial bounds derived in Ref.\,\onlinecite{Kern:2025:BHE} and are much tighter than the bounds that were previously known \cite{Briane:2009:GHE,Briane:2011:BSF}. Furthermore, we will see that two points on these bounds are attained by a microstructure. However, beyond that, there are significant gaps between the bounds and the most extremal microstructures that we were able to identify, indicating that there is room for improvement. We expect that this gap is attributable to the bounds not being tight enough. In fact, we speculate that the microstructures attaining the bounds on the complex effective permittivity \cite{Kern:2020:TBE,Milton:1981:BCP} are optimal for the Hall effect too. This leads us to a set of conjectured bounds, with these microstructures on their boundaries, which would allow for stronger conclusions than the bounds that we derive here, specifically on the corresponding volume-fraction independent bounds and the range of the effective Hall coefficient.

\subsection{The bounds and their attainability}\label{boundsatt}

\begin{figure}[h!]
	\includegraphics[width=\textwidth]{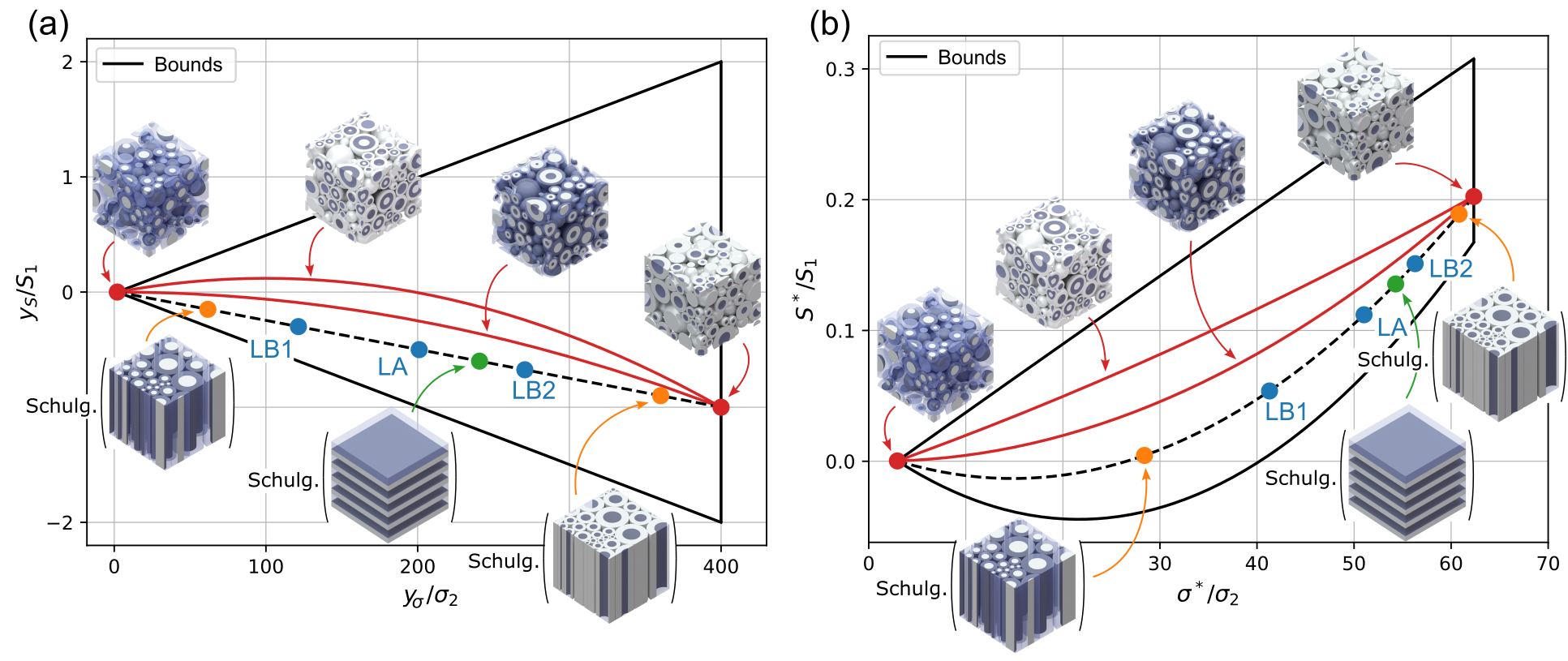}
	\caption{Volume-fraction dependent bounds on the effective Hall parameters of isotropic two-phase metamaterials. Panels (a) and (b) show the bounds in the $y$-plane and the effective parameter plane, respectively. The colored dots and lines correspond to the following microstructures, as illustrated via the insets: The red dots and lines correspond to assemblages of singly- and doubly-coated spheres, respectively. The remaining microstructures are obtained by forming a Schulgasser laminates from a rank-$1$ laminate (green dot), coated cylinder assemblages (orange dots), and the more complicated hierarchical laminates LA1/LA2, LB1, and LB2 (blue dots). Note that the $y$-parameters of these structures lie on a straight line (black/dashed). The outer red curve and the dashed black line are conjectured to correspond to improved bounds. Parameters are $\sigma_1/\sigma_2=200$, $S_2=0$, and $f_1=0.4$.}
	\label{fig1}
\end{figure}

\noindent
The effective $y$-parameters of any isotropic Hall metamaterial made from two isotropic phases are constrained to the isosceles triangle in the $(y_{\sigma},\,y_S)$-plane defined by
\begin{equation}
	|y_S| \leq \frac{y_{\sigma}-2\sigma_2}{\sigma_1-\sigma_2}|S_1B| \text{ and } y_{\sigma} \leq 2\sigma_1,
	\label{eq:boundsyplane}
\end{equation}
where we have assumed that $S_2=0$. Note that the vertex angle and the orientation are the same as in the uniaxial case \cite{Kern:2025:BHE}. The difference to the uniaxial case is that (i) the vertex is shifted to the right and (ii) that the largest value of $y_{\sigma}$, which corresponds to the Hashin-Shtrikman upper bound on the zmf conductivity, is finite, i.e., the region is a triangle rather than a wedge. Once transformed to the effective parameter plane, the bounds form a lens-shaped region that is cut off vertically on the right. Analytic expressions for the bounds in the effective parameter plane are given in the supplementary material. An illustration of the bounds and comparison with the most extreme structures that we were able to identify is shown Fig.\,\ref{fig1}. 

From the bounds (\ref{eq:boundsyplane}), one can obtain corresponding volume-fraction independent bounds by taking the union over all volume-fractions in the effective parameter plane. The resulting volume-fraction independent upper bound is given by the parabola passing through the points corresponding to the two pure phases, i.e., through $(\sigma_2,~0)$ and $(\sigma_1,~S_1)$, and having a minimum at $\sigma^*=\sigma_2/2-\sigma_1^2/\sigma_2$. This result improves slightly over its uniaxial counterpart \cite{Kern:2025:BHE}. The lower bound is the parabola passing through the points corresponding to the two pure phases and having a minimum at 
\begin{equation}
	\sigma^*=\frac{1}{4}\left(\sigma_1-\sqrt{3}\sqrt{\sigma_2(2\sigma_1+\sigma_2)}\right).
\end{equation}
This result implies that an isotropic two-phase composite exhibiting a sign-inversion can only be formed if the conductivity ratio satisfies $\sigma_1 > 13\sigma_2$, which is more restrictive than the corresponding constraint for uniaxial metamaterials \cite{Kern:2025:BHE}, $\sigma_1 > 9\sigma_2$. Note that, while there is an improvement at low and intermediate conductivity ratios, the lower bound becomes identical to its uniaxial counterpart if the conductivity ratio becomes infinitely large.
\smallskip

We now turn to the question of attainability. The left-most vertex of the isosceles triangle is attained by an assemblage of coated spheres. Coated-sphere assemblages are formed by first coating a sphere made from one of the phases with a spherical shell made from the other phase and subsequently assembling scaled copies of this coated sphere such that they fill all space, see Chap.\,7 in Ref.\,\onlinecite{Milton:2002:TOC} for a detailed discussion. Specifically, the left-most vertex is attained by a coated sphere assemblage that uses phase $1$ as the core phase. The $y$-parameters of this microstructure are 
\begin{equation}
	y_{\sigma}^{\text{CS1}} = 2\sigma_2,~y_S^{\text{CS1}}=0.
\end{equation}
The phase-inverted version of this assemblage, i.e., the coated sphere assemblage with phase $2$ as the core phase and phase $1$ as the coating phase, attains the vertical edge of the triangle. Its effective $y$-parameters are given by
\begin{equation}
	y_\sigma^{\text{CS2}} = 2\sigma_1,~y_S^{\text{CS2}}=-S_1B.
\end{equation}
These are the only microstructures attaining the bound for isotropic materials that we were able to identify. The most extremal composites we found are the ones attaining the bounds on the isotropic effective complex permittivity of three-dimensional composites \cite{Milton:1981:BCP,Kern:2020:TBE}. These are, in addition to the coated sphere assemblage and its phase-interchanged version, one of the doubly-coated sphere assemblages and several hierarchical laminates whose effective permittivity has a single pole when expressed as a function of its constituent's permittivities. We refer to Ref.\,\onlinecite{Kern:2020:TBE} regarding the details on how these structures are formed. In the following, we briefly list these microstructures and state their effective properties. 
\smallskip

If one adds a second concentric coating made from the core phase to the spheres, one obtains the two doubly-coated sphere assemblages. Their effective $y$-parameters lie on the circular arcs passing through the two points corresponding to the (singly-)coated sphere assemblages and $(-\sigma_1,~S_1)$ (if phase $1$ is the core phase) or $(-\sigma_2,~0)$ (if phase $2$ is the core phase). The outer one of the circular arcs corresponds to the assemblage with phase $1$ as the core phase.
\smallskip

We now consider the hierarchical laminates having a single pole that were previously shown to have optimal complex permittivities. In the $y$-plane, these laminates lie on a straight line passing through the points corresponding to the coated sphere assemblages. The key idea \cite{Milton:1981:BCP} underlying their conception is to identify an anisotropic microstructure having a single pole when $\varepsilon_2=1$ and the effective complex permittivity is examined as a function of $\varepsilon_1$. Subsequently, this anisotropic structure is made isotropic by forming a Schulgasser laminate \cite{Schulgasser:1977:BCS}. In Ref.\,\onlinecite{Milton:1981:BCP}, Milton introduced Schulgasser laminates formed from the two coated cylinder assemblages and from a rank-$1$ laminate. A Schulgasser laminate formed from a coated cylinder assemblage with phase $1$ as the core phase has effective $y$-parameters
\begin{equation}
	\widetilde{y}^{\text{CC1}} = f_1\widetilde{y}^{\text{CS1}}+f_2\left(\frac{1}{4}\widetilde{y}^{\text{CS2}}+\frac{3}{4}\widetilde{y}^{\text{CS1}}\right),
\end{equation}
while for phase $2$ as the core phase one obtains
\begin{equation}
	\widetilde{y}^{\text{CC2}} = f_2\widetilde{y}^{\text{CS2}}+f_1\left(\frac{1}{4}\widetilde{y}^{\text{CS1}}+\frac{3}{4}\widetilde{y}^{\text{CS2}}\right).
\end{equation}
If the underlying anisotropic microstructure is a rank-$1$ laminate instead, then the properties of the Schulgasser laminate become
\begin{equation}
	\widetilde{y}^{\text{L}} = f_1\widetilde{y}^{\text{CS1}}+f_2\widetilde{y}^{\text{CS2}}
\end{equation}
Note that, as the volume fraction is varied, this point traverses the entirety of the line segment between the two points corresponding to the coated sphere assemblages.
\smallskip

Additional such laminates were introduced in Ref.\,\onlinecite{Kern:2020:TBE}. The first of these laminates is based on the layered coated cylinder assemblages discussed above if the volume fraction of the perpendicular pure layers in the last step is chosen as $(1+f_i)/(4-2f_i)$, with $i=1,\,2$ denoting the core phase. Adopting the notation from Ref.\,\onlinecite{Kern:2020:TBE}, we label this anisotropic laminate as LA1 (if phase $1$ is the core phase) or LA2 (if phase $2$ is the core phase). LA1 can be formed if $f_1$ lies in the interval $\left[1/2,\,1\right]$, whereas LA2 can be formed if $f_1$ lies in the interval $\left[0,\,1/2\right]$. In the $y$-plane, Schulgasser laminates formed from LA1 or LA2 lie at the midpoint between the points corresponding to the coated sphere assemblages, i.e.,
\begin{equation}
	\widetilde{y}^{\text{LA1}} = \widetilde{y}^{\text{LA2}} = \frac{1}{2}\left(\widetilde{y}^{\text{CS1}}+\widetilde{y}^{\text{CS2}}\right).
\end{equation}

The second laminate is obtained by replacing the pure phase in the perpendicular layers by a rank-$1$ laminate, wherein the layers of the rank-$1$ laminate are perpendicular to the cylinder axes and the volume fractions of the core phase~$i$ within the rank-$1$ laminate and within the coated cylinder assemblage are $(2+f_i)/4$ and $f_i/2$, respectively. Note that this implies that the rank-$1$ laminate and the coated cylinder assemblage are combined in fractions $f_{\text{L}}=2f_i/(2-f_i)$ and $1-f_{\text{L}}=(2-3f_i)/(2-f_i)$, respectively, to form the second laminate. Again using the same notation as in Ref.\,\onlinecite{Kern:2020:TBE}, we label the second laminate as LB1 (if phase $1$ is the core phase) or LB2 (if phase $2$ is the core phase). LB1 can be formed if $f_1\in\left[0,\,2/3\right]$, whereas LB2 can be formed if $f_1\in\left[1/3,\,1\right]$. A Schulgasser laminate formed from LB1 has effective $y$-parameters
\begin{align}
	\widetilde{y}^{\text{LB1}}=(1-f_{\text{L}})\widetilde{y}^{\text{CC1}}+f_{\text{L}}\widetilde{y}^{\text{L}}
\end{align}
whereas for a Schulgasser laminate formed from LB2, we obtain
\begin{align}
	\widetilde{y}^{\text{LB2}}=(1-f_{\text{L}})\widetilde{y}^{\text{CC2}}+f_{\text{L}}\widetilde{y}^{\text{L}}.
\end{align}

We speculate that the discussed structures are in fact optimal, i.e., that there is an improved set of bounds corresponding to the straight line passing through the points $(2\sigma_2,\,0)$ and $(2\sigma_1,\,-S_1)$ and the circular arc passing through these points, and, when extended trough $(-\sigma_1,~S_1)$. One indication that this may be the case, is the fact that the corresponding bounds for uniaxial metamaterials are indeed attained by microstructures that also attain the bounds on the effective complex permittivity \cite{Milton:1981:BCP,Kern:2025:BHE}. Furthermore, the effective properties of a wide range of hierarchical laminates that we have generated all fell within these conjectured bounds.
\smallskip

How the conjectured bounds can be proved is presently unclear. One potential avenue is to use the Cherkaev-Gibiansky-Milton transformation \cite{Cherkaev:1994:VPC,Milton:1991:FER,Briane:2011:BSF} in combination with the idea of embedding \cite{Tartar:1985:EFC,Avellaneda:1988:ECP,Milton:1990:BRT}. This avenue can be expected to yield bounds for isotropic metamaterials (applying for arbitrary rather than just for small magnetic fields) but the analysis would likely be involved. Another avenue is to stick to the approach pursued in the present paper but to increase the conductivity tensors of the phases to uniaxial, $\bm{\sigma}_i^+= \text{diag}(\sigma_i+|\eta_i|,\sigma_i+|\eta_i|,\sigma_i)$, rather than to isotropic, $\bm{\sigma}_i^+= (\sigma_i+|\eta_i|)\bm{I}$, tensors. However, whether this strategy yields improved bounds remains to be seen. Note that employing the optimal bound on the complex permittivity proved in Ref.\,\onlinecite{Kern:2020:TBE}, which corresponds to an assemblage of doubly-coated spheres, via an ``inverse Schulgasser-Hashin argument'' \cite{Schulgasser:1976:BEP} only leads to a marginal improvement over our lower bound and is thus not further discussed.
\smallskip

As regards the remaining gaps between the points that lie on the straight line in the $y$-plane, it appears reasonable to expect that the approach in Ref.\,\onlinecite{Kern:2020:TBE} of interpolating between these points by laminating the underlying anisotropic structures (and subsequently forming a Schulgasser laminate) transfers to the nonreciprocal case.  
\smallskip

\begin{figure}[h!]
	\centering
	\includegraphics[width=0.5\textwidth]{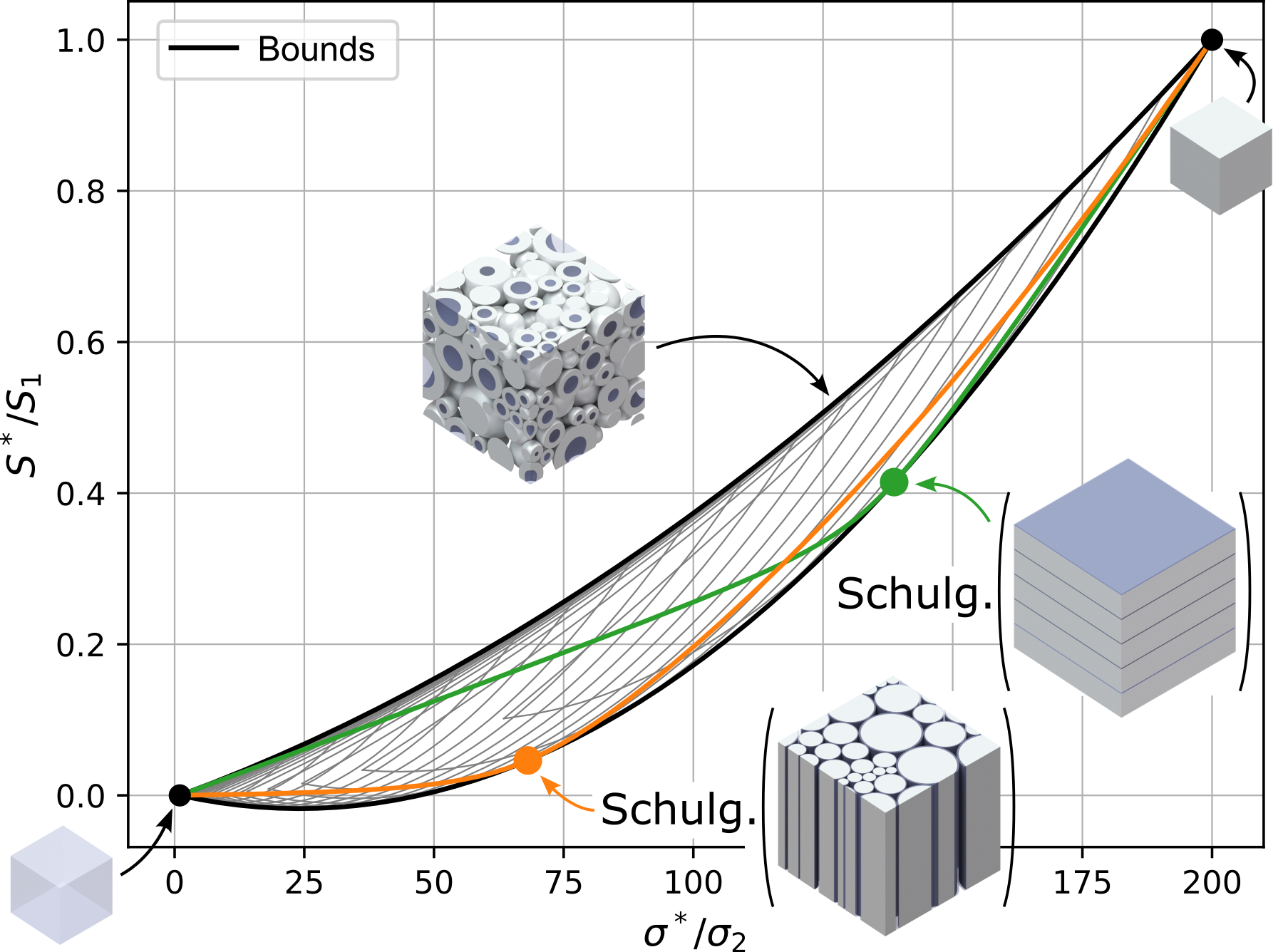}
	\caption{Conjectured volume-fraction independent bounds on the effective Hall coefficient of two-phase isotropic composites. The bounds (black curves) are parabolas passing through the points corresponding to the two pure phases (with phase $1$ shown in gray and phase $2$ in semi-transparent blue). The upper bound is attained by coated sphere assemblages with phase $2$ as the core phase. The green and orange curve correspond to Schulgasser laminates formed from a rank-$1$ laminate and a coated cylinder assemblage, respectively. They attain the conjectured bound at the position marked by the dots for fairly large values of $f_1$ (cf. the insets). The thin gray lines correspond to the volume-fraction dependent bounds for a number of selected volume fractions between zero and one. Parameters are $\sigma_1/\sigma_2=200$, $S_2=0$.}
	\label{fig2}
\end{figure}

If the conjectured bounds were to be proven, then the corresponding volume-fraction independent bounds, which are illustrated in Fig.\,\ref{fig2} would be as follows: The implied upper bound would read
\begin{equation}
	S^* \leq \frac{\sigma^*-\sigma_2}{\sigma_1 - \sigma_2}\frac{\sigma_1 + \sigma_2 + \sigma^*}{2 \sigma_1 + \sigma_2}S_1.
\end{equation}

If established, this bound would be optimal, as it would be attained in its entirety by assemblages of coated spheres with phase $2$ as the core phase. The expression for the implied lower volume-fraction independent bound would depend on the ratio of conductivities of the constituent materials. 
For $\sigma_1\leq4\sigma_2$, the lower bound would be given by
\begin{equation}
	S^* \geq \frac{\sigma^*-\sigma_2}{\sigma_1 - \sigma_2}\frac{2\sigma_2 + \sigma^*}{2 \sigma_2 + \sigma_1}S_1,
\end{equation}
which would be attained in its entirety by coated sphere assemblages with phase $1$ as the core phase. For $\sigma_1>4\sigma_2$, the lower bound would be given by the parabola passing through the points corresponding to the two pure phases and having a minimum at 
\begin{equation}
	\sigma^*=\frac{1}{6}\left(\sigma_1+\sigma_2-4\sqrt{\sigma_1\sigma_2}\right).
\end{equation}

Two points on this lower bound would be attained. The first point corresponds to the Schulgasser laminate formed from a rank-$1$ laminate if the volume fraction is chosen as $f_1=(2\sigma_1-\sqrt{\sigma_1\sigma_2})/(2(\sigma_1-\sigma_2))$. The second point corresponds to the Schulgasser laminate formed from a coated cylinder assemblage if the volume fraction is chosen as $f_1=((\sigma_1+\sigma_2)^2+2(\sqrt{\sigma_1\sigma_2}(\sigma_2-\sigma_1)-2\sigma_2^2)/(\sigma_1-\sigma_2)^2$. For other volume fractions these structures still come close to the conjectured bounds -- in particular for small values of $\sigma_1/\sigma_2$ and with the exception of the sign-inverted regime. Sign-inversions, for which the conjectured bounds would imply the more restrictive condition $\sigma_1 > 25 \sigma_2$, will be treated in further detail in the discussion of the accessible range of the effective Hall coefficient below.

\subsection{Derivation of the bounds}

\noindent
The main idea behind the derivation of the bounds is to use the monotonicity result (\ref{eq:estimatebound}) in conjunction with conventional bounds incorporating known values. The additional ingredient that allowed us to obtain tight bounds in the uniaxial case was the field equation recursion method. Here we recall the corresponding results from Ref.\,\onlinecite{Kern:2025:BHE} and extend the bounds to incorporate the additional information that the metamaterial is isotropic.
\smallskip

In the derivation, the effective properties are expressed through $y$-tensors rather than through effective conductivity tensors. More precisely, we encounter three different $y$-tensors: First, we have the $y$-tensor $\widetilde{\bm{Y}}$ given by (\ref{eq:ytransf}) corresponding to the magnetic-field dependent conductivity tensor. This $y$-tensor has an isotropic part, with coefficient $y_{\sigma}$, and an antisymmetric part, with coefficient $y_S$, cf. Eq.\,(\ref{eq:ytransf}). As we are assuming that the magnetic field is weak, the isotropic part is identical to a second $y$-tensor, $\bm{Y}=y_{\sigma}\bm{I}$, which corresponds to the zero magnetic-field conductivity tensor. Third, consider the fictitious metamaterial that we formed by increasing the conductivities of the phases $\sigma_{1}^+=\sigma_1+|\eta_1|$, $\sigma_{2}^+=\sigma_2+|\eta_2|$. Associated with $\bm{\sigma}^{*+}$ is a corresponding $y$-tensor, $\bm{Y}^+=y^+\bm{I}$, given by an expression analogous to (\ref{eq:ytransf}), i.e.,
\begin{equation}
	y^+ =-f_2\sigma_{1}^+-f_1\sigma_{2}^++\frac{f_1f_2(\sigma_{1}^+-\sigma_{2}^+)^2}{f_1\sigma_{1}^++f_2\sigma_{2}^+-\sigma^{*+}}.
	\label{eq:ytensincr}
\end{equation}
Ultimately, we want to find a bound on the components of $\widetilde{\bm{Y}}$. Recognizing that the magnetic field can be treated as if it were purely imaginary, the idea is to first find an upper bound on $y^+$ and then invoke the monotonicity argument discussed in Sec.\,\ref{derstrategy}. In terms of the $y$-coefficients, the monotonicity result (\ref{eq:estimatebound}) takes the form 
\begin{equation}
	|y_S| \leq y^+-y_{\sigma}.
	\label{eq:monoytens}
\end{equation}
Thus, any upper bound on $y^+-y_{\sigma}$ gives an upper bound on $|y_S|$. Furthermore, the bound should allow us to incorporate the additional information that the metamaterial is isotropic. Here, we use the two-phase bound derived in Ref.\,\onlinecite{Kern:2025:BHE} based on the field-equation recursion method. This bound is given by
\begin{equation}
	y^+-y_{\sigma} \leq |S_1B|\frac{y_{\sigma}-n\sigma_2}{\sigma_1-\sigma_2},
	\label{eq:boundsgen}
\end{equation}
where $n=y^+(1,1)=y_{\sigma}(1,1)$ is a normalization factor that depends on the second derivative of the effective conductivity. Specifically, expanding $\sigma^{*+}$ in orders of $\sigma_1^+-1$ in (\ref{eq:ytensincr}), we obtain
\begin{equation}
	\frac{\partial^2\sigma^{*+}(\sigma_1,1)}{\partial \sigma_1^2}=-\frac{2}{1+n}f_1f_2.
\end{equation}
It is well known that the second derivative of the effective conductivity reflects the symmetry of the metamaterial. For isotropic metamaterials, the effective conductivity satisfies 
\begin{equation}
	\frac{\partial^2\sigma^{*+}(\sigma_1,1)}{\partial \sigma_1^2}=-\frac{2}{3}f_1f_2,
\end{equation}
where $f_1$ and $f_2$ are the volume-fractions of the two phases \cite{Bergman:1978:DCC}. Thus, we obtain $n=2$, which using (\ref{eq:boundsgen}) and (\ref{eq:monoytens}) yields the legs of the triangle (\ref{eq:boundsyplane}), while the (vertically-oriented) base of the triangle corresponds to the upper Hashin-Shtrikman bound on the zmf conductivity \cite{Hashin:1962:VAT}. 

\subsection{Implications for the effective Hall coefficient}

\noindent
In many Hall experiments, the total electric current is prescribed. In this case, the relevant material property is the (effective) Hall coefficient, as it describes the influence of the material on the resulting Hall voltage \cite{Popovic:2003:HED}. Accordingly, much of the previous work on the Hall effect in metamaterials has focused on the effective Hall coefficient. In particular, it has been shown that this quantity can be enhanced \cite{Briane:2009:GHE}, sign-inverted \cite{Briane:2009:HTD,Kadic:2015:HES,Kern:2016:EES}, and both simultaneously \cite{Kern:2019:HBD,Kern:2018:THE}. Naturally, our bounds also constrain the effective Hall coefficient. For example, using the macroscopic version of the relation (\ref{eq:cof}), we can find representations of the bounds in the $(\sigma^*,\,A^*)$-plane. 
\smallskip

\begin{figure}[h!]
	\centering
	\includegraphics[width=0.5\textwidth]{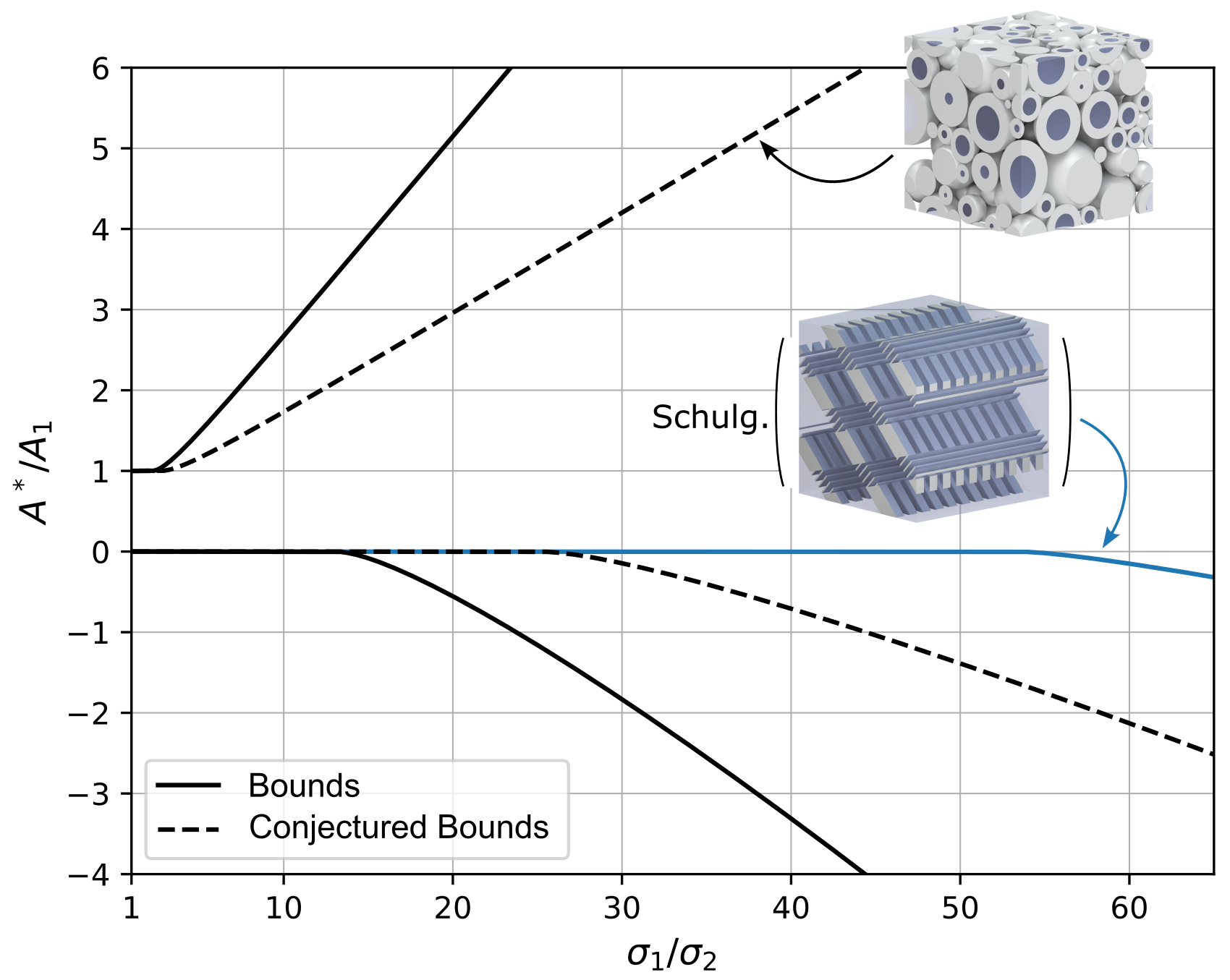}
	\caption{Bounds on the effective Hall coefficient (black lines) of two-phase metamaterials as a function of the conductivity contrast, $\sigma_1/\sigma_2$. Additionally, the conjectured bounds are shown (dashed black lines). The upper conjectured bound is attained by a coated sphere assemblage with phase $2$ as the core phase. The blue curve corresponds to a Schulgasser laminate formed from the anisotropic rank-$3$ laminate introduced in \cite{Kern:2023:SRE}.}
	\label{fig3}
\end{figure}

In the following, we focus on the effective Hall coefficient exclusively. That is, we seek the maximum and minimum of the effective Hall coefficient permitted by our bounds, while allowing the effective conductivity and the volume fraction to vary over all admissible values. The resulting permissible range of the effective Hall coefficient is illustrated in Fig.\,\ref{fig3}. Shown are both the range implied by the bounds (\ref{eq:boundsyplane}) and the narrower range implied the conjectured bounds. Corresponding analytic expressions can be found in the supplementary material. Throughout this section, we assume that $A_2=0$. Again, corresponding results for other values of $A_2$ can be obtained through translations. However, since it is the $\eta$-coefficient that is shifted, the effects are slightly more subtle, cf. Chap.\,4.4 in Ref.\,\onlinecite{Milton:2002:TOC}. 
\smallskip

The permissible range of the effective Hall coefficient depends on the conductivity contrast, $\sigma_1/\sigma_2$, as follows: For low conductivity contrasts, no improvement over the pure phases is possible -- both enhancements and sign-inversions require the contrast to exceed certain thresholds. For enhancements, the conductivity contrast has to satisfy at least $\sigma_1/\sigma_2\geq1+\sqrt{3/2}$, which is slightly more restrictive than the previous result for uniaxial metamaterials, $\sigma_1/\sigma_2\geq2$. If the conjectured bounds were to be proven, this result would further improve to $\sigma_1/\sigma_2\geq1+\sqrt{3}$. Sign-inversions of the effective Hall coefficient occur if and only if the effective $S$-coefficient is sign-inverted, with the corresponding thresholds provided in Sec.\,\ref{boundsatt}. Above these thresholds, the permissible range widens with increasing conductivity contrast, ultimately allowing for arbitrarily large values of the effective Hall coefficient (whether sign-inverted or not) in the limit $\sigma_1/\sigma_2\rightarrow\infty$, as consistent with Refs.\,\onlinecite{Briane:2009:HTD,Briane:2009:GHE,Kadic:2015:HES,Kern:2016:EES,Kern:2018:THE,Kern:2019:HBD}.
\smallskip

Regarding attainability, we note that the conjectured upper bound is realized by coated sphere assemblages with phase $2$ as the core phase. A suitable starting point for evaluating the lower bound is the anisotropic rank-$3$ laminate introduced in \cite{Kern:2023:SRE}. This metamaterial exhibits a sign-inversion of a diagonal component of the effective Hall tensor. This anisotropic behavior can be translated into a sign-inversion of the effective Hall coefficient by forming an isotropic microstructure via Schulgasser's lamination scheme \cite{Schulgasser:1977:BCS,Kern:2023:SRE}. The results of a numerical minimization of the effective Hall coefficient over the geometry parameters of the resulting Schulgasser laminate are shown in Fig.\,\ref{fig3} (see Sec.\,IV of Ref.\,\onlinecite{Kern:2023:SRE} for details). Notably, there remains a substantial gap between the values achieved  achieved by this structure and the conjectured bound. Thus, even if the conjectured bounds were to be proven, the sign-inverted regime would still require further investigation.

\section{Multi-phase volume-fraction dependent bounds}

\noindent
In this section, we provide bounds applying to uniaxial metamaterials made from an arbitrary number of isotropic phases. In contrast to the bounds derived in Ref.\,\onlinecite{Kern:2025:BHE}, the bounds presented here incorporate the volume-fractions of the phases. The derivation uses the monotonicity result (\ref{eq:estimatebound}) in combination with bounds for reciprocal uniaxial metamaterials made from isotropic phases that are based on a result by Prager \cite{Prager:1969:IVB}. 

\subsection{The bounds and their attainability}

\noindent
Assume that we form a uniaxial metamaterial from $m$ isotropic phases with zmf conductivities $\sigma_i$, $S$-coefficients $S_i$, and prescribed volume fractions, $f_i$, $i=1,\dots,m$. Then, the axial effective $S$-coefficient of the metamaterial satisfies
\begin{equation}
	|S_{\parallel}^*|\leq|S_l|\frac{\sigma_{\perp}^*}{\sigma_l}-\frac{(\sigma_{\perp}^*-\sigma_l)^2}{\sigma_l\sum_{i=1, i\neq l}^m f_i(\sigma_i-\sigma_l)^2/(|S_l|\sigma_i-|S_i|\sigma_l)},
	\label{eq:multiphasebound}
\end{equation}
where the $l$-th constituent material has the unique largest mobility among the constituent materials, i.e., $|S_i|/\sigma_i<|S_l|/\sigma_l$ for all $i\neq l$. If two or more phases share the largest mobility, then the bound reduces to the straight-line bound $|S^*|/\sigma_{\perp}^*\leq|S_l|/\sigma_l$.
\smallskip

\begin{figure}[h!]
	\centering
	\includegraphics[width=\textwidth]{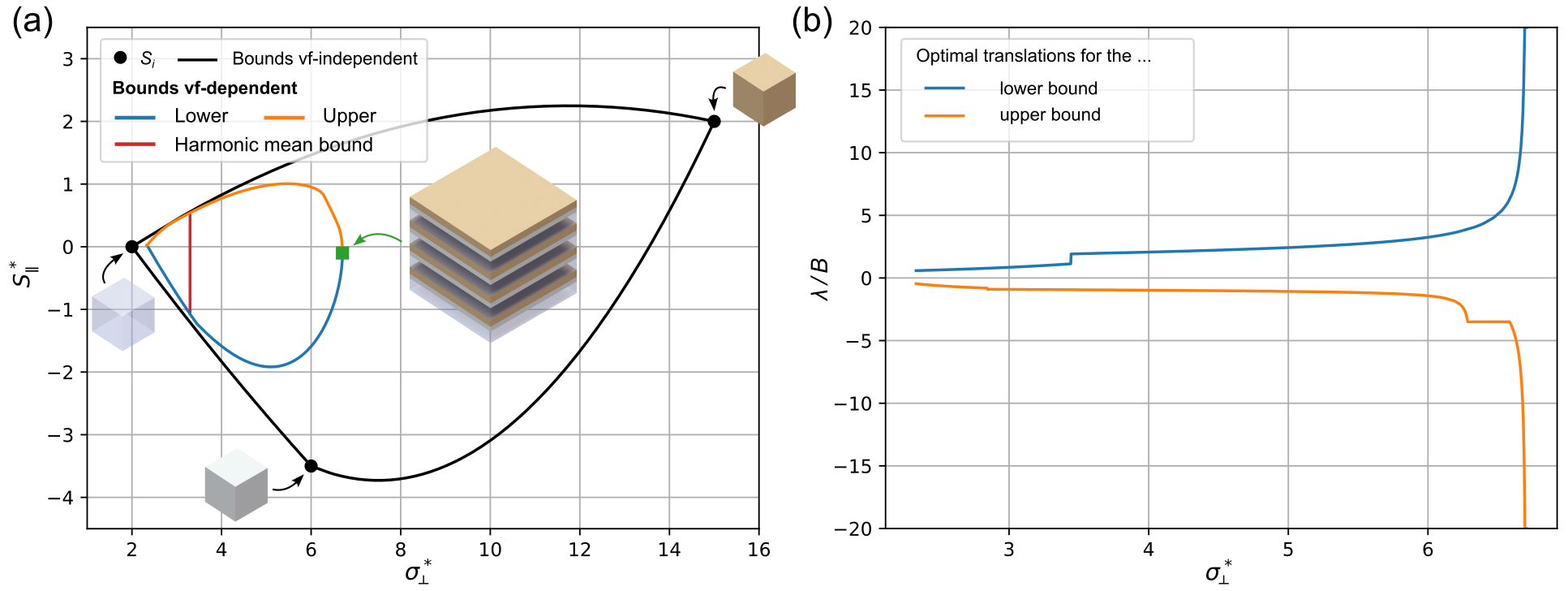}
	\caption{(a) Illustration of the volume-fraction dependent bounds (orange and blue curves) on the effective axial $S$-coefficient of multi-phase metamaterials as a function of the transverse effective conductivity. The right endpoint of the bounds (green square) is attained by a rank-$1$ laminate. The bounds are further tightened through the harmonic mean bound (red line). For reference, the volume-fraction independent bounds (black curves) introduced in Ref.\,\onlinecite{Kern:2025:BHE} are provided. The translations for the volume-fraction independent bounds are (i) $\lambda=\eta_3$, (ii) $\lambda$ chosen such that all $|\eta_i-\lambda|$ lie on the same parabolic-arc bound, and (iii) $\lambda$ chosen such that $|\eta_1-\lambda|/\sigma_1=|\eta_2-\lambda|/\sigma_2$. For the volume-fraction dependent bounds, we have determined the optimal value of $\lambda$, i.e., the value resulting in the tightest bound, numerically and separately for each value of $\sigma_{\perp}^*$. The resulting optimal values of $\lambda$, corresponding to the bounds in panel (a), are shown in panel (b). Note that the absolute value of $\lambda/\eta$ diverges at the right endpoint. Parameters are $\sigma=(2,\,6,\,15)$, $S=(0,\,-3.5,\,3)$, and $f_1=0.5$, $f_2=0.2$.}
	\label{fig4}
\end{figure}

As in many other instances, translations are the key to obtain tightened bounds. As laid out in Sec.\,\ref{derstrategy}, we shift the antisymmetric part of the conductivity tensor, i.e., the $\eta$-coefficients, by a translation parameter, $\lambda$, i.e., $\eta_i \rightarrow \eta_i -\lambda$, which causes the effective $\eta$-coefficient to shift in the same way, $\eta^* \rightarrow \eta^* -\lambda$. One example of the resulting bounds is illustrated in Fig.\,\ref{fig4}. Note, in particular, that the bounds are much tighter than the previously derived volume-fraction independent bounds \cite{Kern:2025:BHE}, at least for this choice of parameters. In addition to the bounds derived here, one should consider the harmonic mean bound on the zmf conductivity, as this further restricts the range  of the effective parameters, i.e., even after translations, the bounds (\ref{eq:multiphasebound}) do not inherently imply the constraint imposed by the harmonic mean bound. 
\smallskip

Irrespective of the chosen parameters, one point on the bounds is attained by a rank-$1$ laminate. Specifically, the rank-$1$ laminate attains the right endpoint of the bounds, where they coincide with the arithmetic mean bound on the transverse effective zmf conductivity. To see this, we first note that the relevant effective Hall properties of a rank-$1$ laminate are given by
\begin{equation}
	\sigma_{\perp}^* = \sum_i^m f_i\sigma_i,~ S_{\parallel}^* = \sum_i^m f_iS_i.
	\label{eq:proprankone}
\end{equation} 
Next, we perform a translation and take the limit $\lambda/\eta\rightarrow\infty$ in the bound (\ref{eq:multiphasebound}), which yields 
\begin{equation}
	0 \geq (\sigma_l-\sigma_{\perp}^*)\left(1+\frac{\sigma_l-\sigma_{\perp}^*}{\sum_{i=1}^{m}f_i\sigma_i-\sigma_l}\right).
	\label{eq:boundzero}
\end{equation} 
Since, in the limit $\lambda/\eta\rightarrow\infty$, $l$ corresponds to the phase with the smallest conductivity, Eq.\,(\ref{eq:boundzero}) yields $\sigma_{\perp}^* = \sum_i^m f_i\sigma_i$, confirming that the bound coincides with the arithmetic mean bound and that it is attained by the rank-$1$ laminate. 
\smallskip

The bounds also uniquely determine the corresponding effective axial $S$-coefficient. More precisely, if the effective transverse conductivity of a uniaxial metamaterial is given by the arithmetic mean as in Eq.\,(\ref{eq:proprankone}), then so is the effective axial $S$-coefficient. This follows from examining the terms of order $\eta/\lambda$ in an expansion of the translated bounds around $\lambda/\eta\rightarrow\infty$, which yields $S_{\parallel}^* \geq \sum_i^m f_iS_i$, and by performing the corresponding analysis for $\lambda/\eta\rightarrow-\infty$, which yields $S_{\parallel}^* \leq \sum_i^m f_iS_i$, together establishing  $S_{\parallel}^* = \sum_i^m f_iS_i$.
\smallskip

In general, however, it appears that the bounds may be further improved. We infer this by comparing the multi-phase bounds (specialized to $m=2$) with the two-phase bounds for uniaxial metamaterials previously derived in Ref.\,\onlinecite{Kern:2025:BHE} (not depicted). In the examples that we investigated, the previously derived lower bound (assuming $S_1>0$) was significantly tighter. Presumably, this suboptimality persists for $m>2$.
\smallskip

Finally, note that the uniaxial bounds can be extended to fully anisotropic metamaterials. The idea is to form a fictitious uniaxial metamaterial from the fully anisotropic metamaterial. Then, one can apply the uniaxial bounds to the former, which implies bounds on the latter. Corresponding details are discussed in Ref.\,\onlinecite{Kern:2025:BHE}.

\subsection{Derivation of the bounds}

\noindent
In Ref.\,\onlinecite{Prager:1969:IVB}, Prager derived bounds for the effective conductivity of metamaterials incorporating that a certain known microscopic conductivity results in a certain known effective conductivity. In the following, we sketch and generalize Prager's derivation before using the result to obtain multi-phase bounds for the effective Hall parameters of uniaxial metamaterials. Note that our notation, specifically the use of the superscript ``+'', is flipped as compared to Prager's: We aim to derive bounds on $\bm{\sigma}^{*+}$ (the effective conductivity tensor of the fictitious reciprocal metamaterial) while incorporating that a microscopic conductivity $\sigma(\bm{x})$ (the microscopic zero magnetic-field conductivity) results in an effective conductivity $\sigma^*$ (the effective zero magnetic-field conductivity). In contrast to the derivation in Ref.\,\onlinecite{Prager:1969:IVB}, we allow $\bm{\sigma}^{*+}$ to be anisotropic and subsequently consider the uniaxial case. Furthermore, whereas Prager quickly specializes to the two-phase case, we take the multi-phase analysis further. We expect that these extensions of Prager's bounds will be useful in other contexts beyond the Hall effect.
\medskip

\noindent
We start from the elementary variational inequality,
\begin{equation}
	\langle \bm{\underline{e}} \rangle\cdot\bm{\sigma}^{*+} \langle \bm{\underline{e}}\rangle \leq \left\langle \sigma^+ \bm{\underline{e}}^2\right\rangle,
	\label{eq:genvarineq}
\end{equation}
where $\bm{\underline{e}}$ is a trial field. Furthermore, we know that a microscopic conductivity $\sigma(\bm{x})$ gives rise to an effective conductivity $\sigma^*$,
\begin{equation}
	\sigma^{*} \langle \bm{e}\rangle^2 = \left\langle \sigma \bm{e}^2\right\rangle,
	\label{eq:trialvar}
\end{equation}
where $\bm{e}(\bm{x})$ is the electric field solving the corresponding conductivity equations. Prager continues by writing the microscopic conductivity as $\sigma^+=\sigma-(\sigma-\sigma^+)$ and chooses $\bm{e}(\bm{x})$ as the trial field, which yields
\begin{equation}
	\langle \bm{e} \rangle\cdot\bm{\sigma}^{*+} \langle \bm{e}\rangle \leq \sigma^{*} \langle \bm{e}\rangle^2-\left\langle(\sigma-\sigma^+)\bm{e}^2\right\rangle.
\end{equation}
Assuming that $\sigma\geq\sigma^+$ (which generally does not hold and, thus, requires some more attention below) and using the Cauchy-Schwarz inequality, one further obtains, cf. Eq.\,(13b)  in Ref.\,\onlinecite{Prager:1969:IVB},
\begin{equation}
	\sigma^{*} \langle \bm{e}\rangle^2-\langle \bm{e} \rangle\cdot\bm{\sigma}^{*+} \langle \bm{e}\rangle \geq  \frac{\langle(\sigma - \sigma^+)\mathbf{e}\cdot\bm{q}\rangle^2}{\langle(\sigma - \sigma^+)\bm{q}^2\rangle},
	\label{eq:boundani}
\end{equation}
where the vector-valued function $\bm{q}(\bm{x})$ can be freely chosen. We now assume that the fictitious metamaterial is uniaxial, choose $\langle \bm{e}\rangle$ to be normalized and to lie in the transverse plane, and choose $\bm{q}(\bm{x})$ to be collinear with $\bm{e}(\bm{x})$, thereby obtaining
\begin{equation}
	\sigma^{*} - \sigma_{\perp}^{*+} \geq  \frac{\langle(\sigma - \sigma^+)eq\rangle^2}{\langle(\sigma - \sigma^+)q^2\rangle},
	\label{eq:intermresbound}
\end{equation}
where $q=|\bm{q}|$ and $e=|\bm{e}|$. Going further beyond the derivation in Ref.\,\onlinecite{Prager:1969:IVB}, we continue the multi-phase analysis and take $q(\bm{x})$ to be constant in each phase,
\begin{equation}
	q(\bm{x}) = \sum_{i=1}^m \gamma_i\chi_i(\bm{x})/(\sigma_i - \sigma_i^+),
	\label{eq:funcqu}
\end{equation}
where the $\chi_i(\bm{x})$ are the characteristic functions taking the value one in phase $i$ and zero elsewhere. Note that the additional factor $1/(\sigma_i - \sigma_i^+)$ is introduced without loss of generality and merely serves to simplify subsequent formulae. Further, choosing $\gamma_i=\sigma_i+c$, where $c$ is a constant, to ensure that the numerator in Eq.\,(\ref{eq:intermresbound}) only depends on $\sigma^*$, gives 
\begin{equation}
	\left\langle(\sigma - \sigma^+)eq\right\rangle^2 = \left\langle\sum_{i=1}^m \gamma_i\chi_ie\right\rangle^2=(\langle\sigma e\rangle+c)^2=(\sigma^*+c)^2,
	\label{eq:boundnumer}
\end{equation}
while evaluating the denominator in Eq.\,(\ref{eq:intermresbound}) yields
\begin{equation}
	\langle(\sigma - \sigma^+)q^2\rangle = \left\langle\sum_{i=1}^m \frac{\gamma_i^2\chi_i(\bm{x})}{\sigma_i - \sigma_i^+}\right\rangle = \sum_{i=1}^m f_i\frac{(\sigma_i+c)^2}{\sigma_i - \sigma_i^+}.
	\label{eq:bounddenum}
\end{equation}
We now make the replacement $\sigma_i\rightarrow k\sigma_i$, which, due to homogeneity, implies $\sigma^*\rightarrow k\sigma^*$, and obtain from Eqs.\,(\ref{eq:intermresbound}), (\ref{eq:boundnumer}), and (\ref{eq:bounddenum})
\begin{equation}
	k\sigma^{*} - \sigma_{\perp}^{*+} \geq \frac{(k\sigma^{*} + c)^2}{\sum_{i=1}^m f_i(k\sigma_i + c)^2/(k\sigma_i - \sigma_i^+)},
	\label{eq:finbound}
\end{equation}
wherein the optimal value of the constant $c$ is given by
\begin{equation}
	c = -\frac{\sum_{i=1}^m f_ik\sigma_i(\sigma^* - \sigma_i)/(k\sigma_i - \sigma_i^+)}{\sum_{i=1}^m f_i(\sigma^* - \sigma_i)/(k\sigma_i - \sigma_i^+)}.
	\label{eq:finc}
\end{equation}
Under the replacement $\sigma_i\rightarrow k\sigma_i$, the condition $\sigma_i\geq \sigma^+$, which is generally not satisfied, becomes $k\sigma_i\geq \sigma^+$. Thus, we can choose $k$ such that the latter condition holds true. To obtain a tight bound, we want to keep $k$ as small as possible. Setting $\sigma_i^+=\sigma_i+|\eta_i|$, we thus choose 
\begin{equation}
	k=1+\text{max}_i\frac{|\eta_i|}{\sigma_i}+\varepsilon=1+\frac{|\eta_l|}{\sigma_l}+\varepsilon.
	\label{eq:kexp}
\end{equation}
Letting $\varepsilon \rightarrow 0$ and employing the monotonicity result (\ref{eq:estimatebound}), Eqs.\,(\ref{eq:finbound}), (\ref{eq:finc}), and (\ref{eq:kexp}) then yields the bound (\ref{eq:multiphasebound}).

\section{Conclusion}

\noindent
In this paper, we have extended a previous monotonicity-based approach for bounding nonreciprocal effective properties by incorporating additional microstructural information.
\smallskip

For two-phase isotropic metamaterials, we derived both volume-fraction dependent and independent bounds. These new bounds are generally much tighter than those that were previously derived without the monotonicity-based approach. Furthermore, we showed that the bounds imply that a conductivity contrast of $\sigma_1 > 13\sigma_2$ is necessary for an isotropic two-phase metamaterial to exhibit a sign-inversion of the effective Hall coefficient, which improves upon the corresponding constraint for uniaxial metamaterials, $\sigma_1 > 9\sigma_2$. Nevertheless, a comparison with hierarchical laminates revealed that the bounds can presumably be tightened further. This observation led us to conjecture improved bounds corresponding to microstructures that attain known complex permittivity bounds. 
\smallskip

For multi-phase metamaterials, we presented bounds that, in contrast to those derived in Ref.\,\onlinecite{Kern:2025:BHE}, incorporate the volume-fractions of the phases. As with many previous bounds, translations proved to be a key tool to tighten them. When the volume fractions are known, the new bounds generally allow one to constrain the effective Hall parameters much more narrowly than their volume-fraction independent counterparts. Furthermore, for at least one point, the new bounds are optimal, as that point is attained by a rank-$1$ laminate. However, there seems to be substantial room for further improving the bounds, as we inferred from a comparison with previously derived two-phase bounds. 
\smallskip

Collectively, these results refine our picture of the limits of nonreciprocal metamaterials. For future work, a particularly interesting question is whether the conjectured bounds for two-phase metamaterials can be proved, potentially via one of the approaches that we have outlined. 
\medskip

\noindent
\textbf{Acknowledgements} \par 
\noindent
The authors are grateful to the National Science Foundation for support through the Research Grant No. DMS-2107926. C. Kern acknowledges financial support from the Villum Foundation through the Villum Investigator Project Amstrad (VIL54487).

\medskip

\ifx \bblindex \undefined \def \bblindex #1{} \fi\ifx \bbljournal \undefined
\def \bbljournal #1{{\em #1}\index{#1@{\em #1}}} \fi\ifx \bblnumber
\undefined \def \bblnumber #1{{\bf #1}} \fi\ifx \bblvolume \undefined \def
\bblvolume #1{{\bf #1}} \fi\ifx \noopsort \undefined \def \noopsort #1{}
\fi\ifx \bblindex \undefined \def \bblindex #1{} \fi\ifx \bbljournal
\undefined \def \bbljournal #1{{\em #1}\index{#1@{\em #1}}} \fi\ifx
\bblnumber \undefined \def \bblnumber #1{{\bf #1}} \fi\ifx \bblvolume
\undefined \def \bblvolume #1{{\bf #1}} \fi\ifx \noopsort \undefined \def
\noopsort #1{} \fi

\clearpage

\setcounter{page}{1}

\renewcommand{\thesection}{S\Roman{section}}
\renewcommand{\thesubsection}{\Alph{subsection}}

\titleformat{\section}
{\normalfont\large\bfseries}
{\thesection.}
{1em}
{}

\titleformat{\subsection}
{\normalfont\normalsize\bfseries}
{\thesubsection.}
{0.8em}
{}

\setcounter{section}{0}
\setcounter{subsection}{0}

\begin{center}
	
	{\large\textbf{Supplementary Material:} Further fundamental bounds on\\ the Hall effect in three-dimensional metamaterials}
	
	\vspace{1.5em}
	
	{\normalsize
		Christian Kern$^{1}$ and Graeme W. Milton$^{2}$\\
	}
	\vspace{1em}
	
	{\small\it
		$^{1}$Department of Civil and Mechanical Engineering,\\ Technical University of Denmark, 2800 Kgs. Lyngby, Denmark\\
		$^{2}$Department of Mathematics, University of Utah, Salt Lake City, UT 84112, USA
	}
	
\end{center}

\begin{center}
	Email Addresses: physics@chrkern.de, graeme.milton@utah.edu
\end{center}

\vspace{2em}

\makeatletter
\newcommand{\sicontents}{
	\@starttoc{sitoc}
}
\newcommand{\siaddcontentsline}[3]{
	\addcontentsline{sitoc}{#1}{#2}
}
\makeatother
\let\oldsection\section
\let\oldsubsection\subsection
\renewcommand{\section}[1]{
	\oldsection{#1}
	\siaddcontentsline{section}{\makebox[2.5em][l]{\thesection.}#1}{}
}
\renewcommand{\subsection}[1]{
	\oldsubsection{#1}
	\siaddcontentsline{subsection}{\hspace{2.0em}\makebox[2.5em][l]{\thesubsection.}#1}{}
}

\begin{center}
	\noindent\large\textbf{Contents}
\end{center}
\sicontents

\newpage

\section{Analytic expressions -- effective parameter plane}

\noindent
In this section, we provide analytic expressions for the volume-fraction dependent two-phase bounds for isotropic metamaterials in the effective parameter plane (Subsection~\ref{analyticiso}) as well as analytic expressions for the corresponding conjectured bounds (Subsection~\ref{analyticconjiso}).\\

\noindent
Recall that we consider isotropic metamaterials formed from two isotropic phases with conductivities $\sigma_1$, $\sigma_2<\sigma_1$ and $S$-coefficients $S_1>0$, $S_2=0$. For $S_1 < 0$, the upper and lower bound flip roles. Expressions for $S_2 \neq 0$ can be derived readily via a translation.

\subsection{Isotropic bounds}\label{analyticiso}

\noindent
As discussed in Sec. 3\,(a) of the main text, the effective properties of isotropic metamaterials are confined to the isosceles triangle defined by Eq.\,(3.3), which we repeat here,
\begin{equation}
	|y_S| \leq \frac{y_{\sigma}-2\sigma_2}{\sigma_1-\sigma_2}|S_1B| \text{ and } y_{\sigma} \leq 2\sigma_1.
\end{equation}
Via the inverse $y$-transformation, which is a fractional linear transformation and, thus, maps circles or straight lines to circles or straight lines, we obtain corresponding expressions in the effective parameter plane, i.e., in the $(\sigma^*,\,S^*)$-plane. In this plane, the effective $S$-coefficient is confined to a lens-shaped region defined by an upper and a lower bound and cut-off on the right side by a vertical line that corresponds to the Hashin-Shtrikman upper bound on the zmf conductivity,
\begin{equation}
	S_{\text{vf}}^{\text{lower}}(\sigma^*,f_1) \leq S^* \leq S_{\text{vf}}^{\text{upper}}(\sigma^*,f_1) \text{ and } \sigma^*\leq\sigma_1-\frac{3f_2\sigma_1(\sigma_1-\sigma_2)}{3\sigma_1-f_1(\sigma_1-\sigma_2)},
	\label{eq:boundsintro}
\end{equation}
wherein the upper and lower bound are given by 
\begin{align}
	S_{\text{vf}}^{\text{upper}} &= \frac{f_1(\sigma_1-\sigma_2)(\sigma_1+5\sigma_2)(\sigma^*-\sigma_2)-3\sigma_2(\sigma_2-\sigma^*)^2-f_1^2 (\sigma_1 - \sigma_2)^2 (2\sigma_2 +\sigma^*)}{f_1f_2 (\sigma_1 - \sigma_2)^3}S_1, \\
	S_{\text{vf}}^{\text{lower}} &= \frac{f_1(\sigma_1-\sigma_2)(\sigma_2-\sigma^*)(\sigma_1+3\sigma_2+2\sigma^*)+2(\sigma_1+\sigma_2)(\sigma_2-\sigma^*)^2+f_1^2(\sigma_1-\sigma_2)^2(2\sigma_2+\sigma^*)}{f_1f_2 (\sigma_1 - \sigma_2)^3}S_1.
\end{align}

\subsection{Conjectured isotropic bounds}\label{analyticconjiso}

\noindent
In the $y$-plane, the conjectured bounds for isotropic metamaterials are given by a straight line passing through the points $(2\sigma_1,\,-S_1)$ and $(2\sigma_2,\,0)$ and the circular arcs passing through these two points and, when extended, through $(-\sigma_1,-S_1)$. Under our assumptions, the outer circular arc is always the one passing through $(-\sigma_1,-S_1)$.\\

\noindent
In the $(\sigma^*,\,S^*)$-plane, one obtains the following expressions for these bounds
\begin{align}
	S_{\text{vf}}^{\text{upper}} &= \frac{\sigma_2(2f_2(\sigma_1-\sigma_2)(\sigma_1-2\sigma_2)-3\sigma_1^2)+\sigma_1(f_2(\sigma_1-\sigma_2)+6\sigma_2)\sigma^*+(f_2(\sigma_1-\sigma_2)-3\sigma_2)(\sigma^*)^2}{2f_2 (\sigma_1 - \sigma_2)^2(\sigma_1+2\sigma_2)}S_1, \\
	S_{\text{vf}}^{\text{lower}} &= \frac{f_1(\sigma_1-\sigma_2)(\sigma_2-\sigma^*)(\sigma_1+2\sigma_2+3\sigma^*)+3\sigma_1(\sigma_2-\sigma^*)^2+f_1^2(\sigma_1-\sigma_2)^2(2\sigma_2+\sigma^*)}{2f_1f_2 (\sigma_1 - \sigma_2)^3}S_1.
\end{align}
\newpage

\section{Analytic expressions -- bounds on the effective Hall coefficient}

\noindent
In this section, we provide analytic expressions for the bounds on the effective Hall coefficient discussed in Sec.\,3\,(c) of the main text. These expressions follow from the volume-fraction independent bounds on the effective $S$-coefficient as a function of the effective conductivity. In addition to expressions for the proven bounds (Subsection\,\ref{Hallcoeffisobound}), we provide expressions for the conjectured bounds (Subsection\,\ref{Hallcoeffconjbound}).\\

\noindent
Recall that we are considering isotropic metamaterials formed from two isotropic phases with conductivities $\sigma_1$ and $\sigma_2<\sigma_1$ and Hall coefficients $A_1\neq 0$ and $A_2=0$. Results for $A_2\neq0$ can, in principle, be obtained via a suitable translation. This is, however, less straightforward than for the $S$-coefficient, as the antisymmetric part of the conductivity tensor rather than of the resistivity tensor is being shifted.

\subsection{Isotropic bounds}\label{Hallcoeffisobound}

\noindent
In Sec.\,3\,(a) of the main text, we saw that the effective $S$-coefficient is bounded by two parabolas passing through the points corresponding to the pure phases, $(\sigma_2,~0)$ and $(\sigma_1,~S_1)$. The upper and lower parabola have a minimum at
\begin{equation}
	\sigma^*=\frac{\sigma_2}{2}-\frac{\sigma_1^2}{\sigma_2} \text{ and }\sigma^*=\frac{1}{4}\left(\sigma_1-\sqrt{3}\sqrt{\sigma_2(2\sigma_1+\sigma_2)}\right),
\end{equation}
respectively. These bounds imply that the (relative) effective Hall coefficient, $A^*/A_1$, has to lie between an upper and a lower bound with the corresponding analytic expressions given by 
\begin{align}
	A^{\text{upper}} &= \left\{
	\begin{array}{ll}
		1 & \text{if }\sigma_1 \leq (1+\sqrt{3/2}) \sigma_2 \\
		\frac{(2\sigma_1^2+\sigma_2^2)^2}{8\sigma_1\sigma_2(2\sigma_1^2-\sigma_1\sigma_2-\sigma_2^2)} & \text{if }\sigma_1 > (1+\sqrt{3/2}) \sigma_2. \\
	\end{array}\right.\\
	A^{\text{lower}} &= \left\{
	\begin{array}{ll}
		0 & \text{if }\sigma_1 \leq 13 \sigma_2 \\
		\frac{\sigma_1^2 (\sigma_1^2 - 6 \sigma_1 \sigma_2 - 37 \sigma_2^2+2\sqrt{3}\sqrt{\sigma_2 (2 \sigma_1 + \sigma_2)} (10 \sigma_2-\sigma_1))}{4 \sigma_2 (\sigma_2-\sigma_1) (\sigma_1^2 - 10 \sigma_1 \sigma_2 + \sigma_2^2)} & \text{if }\sigma_1 > 13 \sigma_2. \\
	\end{array}\right.~~~~~
\end{align}

\subsection{Conjectured isotropic bounds}\label{Hallcoeffconjbound}

\noindent
In Sec.\,3\,(a) of the main text, we conjectured that the effective $S$-coefficient satisfies the improved (volume-fraction independent) bounds
\begin{equation}
	\frac{\sigma^*-\sigma_2}{\sigma_1 - \sigma_2}\frac{2\sigma_2 + \sigma^*}{2 \sigma_2 + \sigma_1} \leq \frac{S^*}{S_1} \leq \frac{\sigma^*-\sigma_2}{\sigma_1 - \sigma_2}\frac{\sigma_1 + \sigma_2 + \sigma^*}{2 \sigma_1 + \sigma_2}.
\end{equation}
This yields the following expressions for the conjectured bounds on the (relative) effective Hall coefficient, $A^*/A_1$,
\begin{align}
	A^{\text{upper}} &= \left\{
	\begin{array}{ll}
		1 & \text{if }\sigma_1 \leq \left(1+\sqrt{3}\right) \sigma_2 \\
		\frac{\sigma_1^2(\sigma_1+2\sigma_2)^2}{4\sigma_2(\sigma_1-\sigma_2)(\sigma_1+\sigma_2)(2\sigma_1+\sigma_2)} & \text{if }\sigma_1 > \left(1+\sqrt{3}\right) \sigma_2. \\
	\end{array}\right.\\
	A^{\text{lower}} &= \left\{
	\begin{array}{ll}
		0 & \text{if }\sigma_1 \leq 25 \sigma_2 \\
		\frac{\sigma_1^2(\sqrt{\sigma_1}-5\sqrt{\sigma_2})^2 (\sigma_1 - 2 \sigma_2 + 4 \sqrt{\sigma_1 \sigma_2})}{8 \sigma_2 (\sigma_2-\sigma_1) (\sigma_1^2 - 20 \sigma_1 \sigma_2 + 4 \sigma_2^2)} & \text{if }\sigma_1 > 25 \sigma_2. \\
	\end{array}\right.~~~~~
\end{align}

\end{document}